\documentclass[12pt]{iopart}
\usepackage{graphicx}
\usepackage{bm}
\usepackage{amssymb}
\usepackage{cite}

\begin{document}

\title[Is it possible to grow amorphous normal nanosprings~?]
{Is it possible to grow amorphous normal nanosprings~?}

\author{Alexandre F. Fonseca$^1$\footnote{Present address:
Alan McDiarmid Nanotech Institute, University of Texas at Dallas,
Richardson, TX 75083-0688, USA.}, C. P. Malta$^1$ and D. S.
Galv\~{a}o$^2$}

\address{$^1$ Instituto de F\'{\i}sica, Universidade de S\~ao
Paulo, Caixa Postal 66318, 05315-970, S\~ao Paulo, Brazil}

\address{$^2$ Instituto de F\'{\i}sica `Gleb Wataghin',
Universidade Estadual de Campinas, Unicamp 13083-970, Campinas, SP,
Brazil}

\eads{\mailto{afonseca@if.usp.br}, \mailto{coraci@if.usp.br},
\mailto{galvao@ifi.unicamp.br}}

\begin{abstract} 

  Nanosprings have been object of intense investigations in recent
  years. They can be classified as {\it normal} or {\it binormal}
  depending on the geometry of their cross-section. {\it Normal}
  amorphous nanosprings have not been observed experimentally up to
  now so we have decided to investigate into this matter. We discuss
  the shape of the catalyst in terms of the cross-sectional shape of
  the nanospring, and show that, within the vapor-liquid-solid model,
  the growth of amorphous {\it binormal} nanosprings is energetically
  favoured.

\end{abstract} 



\submitto{\NT}

\maketitle

\section{Introduction}  

There exist various methods and processes for growing nanostructures.
Different types of nanostructures require different methods and
processes to be grown. Several groups are working on the development
and improvement of the growth processes motivated not only by new
physical phenomena but also by the great variety of technological
applications~\cite{iijima,baugh,ratner,samu,ede,wo1,gu1,liu,fonseca}.

In particular, the existence of helically shaped nanowires
(nanosprings or nanohelices) is of great interest because of the
potential applications in nanoelectronics, nanomechanics and
nanoelectromechanical systems~\cite{mc1}. Examples of such structures
are quasi-nanosprings~\cite{tang}, helical crystalline
nanowires~\cite{zhangw,ame,wang,wang2}, and amorphous
nanosprings~\cite{mc2,zhang,mc3,mukho,mc4,veri}.

In contrast to the formation of straight nanowires, the synthesis of
helical nanostructures requires either the existence of anisotropy at
some level of the growth process, or the existence of external forces
holding the nanowire into a helical shape. Both cases have been
reported in the literature. In the case of amorphous nanosprings,
McIlroy {\it et al}~\cite{mc1,mc2} have shown, based on the
vapor-liquid-solid (VLS) growth model~\cite{ellis}, that the
anisotropy in the contact angle between the catalyst and the nanowire
induces helical growth. In the case of crystalline nanosprings, Kong
and Wang~\cite{wang} reported the formation of nanohelices of zinc
oxide~(ZnO) and showed that the eletrostatic interaction between the
nanowire, and the substrate where it is grown, holds the ZnO nanowires
in a helical shape.

A helical structure is classified as {\it normal}/{\it binormal}
depending on the orientation of its cross-section with respect to the
{\it normal} or {\it binormal} vectors~\cite{alain3,fonseca1}. We have
analyzed nanosprings of various materials, reported in the
literature~\cite{wang,mc2,zhang,mc3,wang2}, and have not found a
single case of {\it normal} amorphous nanohelices. In the case of
crystalline helical nanostructure, Gao {\it et al}~\cite{gao} have
recently reported the synthesis of a ZnO {\it normal} nanohelix. Why
normal amorphous nanosprings have not been observed~? Using the VLS
model we provide the first theoretical explanation for the
non-existence of amorphous {\it normal} nanohelices. We have extended
the VLS growth model~\cite{mc1,mc2}, so as to take into account
possible asymmetries in the shape of the catalytic particle. We show
that the growth of amorphous {\it binormal} nanosprings is
energetically more favoured in comparison to the growth of {\it
normal} ones.

{\it Normal} and {\it binormal} helical nanostructures may lead to
different technological applications. We have shown that two
nanosprings of same radius and pitch, same material, possessing the
same cross-section geometry, but differing by the fact that one is a
{\it normal} helical structure and the other is {\it binormal}, have
different stiffness~\cite{fonseca1,fonseca2}. In this comparison, the
{\it normal} nanospring is always stiffer than the {\it binormal}
one~\cite{fonseca1,fonseca2}.

In Section 2, we briefly describe the geometry of a helical structure
and give the definition of {\it normal} and {\it binormal} helices.
The reported experimental results are classified according to this
definition. In Section 3 we analyse the shape of the nanospring
cross-sections and discuss the possible shapes of the catalytic
particle necessary to drive the growth of amorphous nanohelices with
non-circular cross-section, through the VLS mechanism. In Section 4,
based on the VLS growth model, we show that the growth of amorphous
{\it binormal} nanospring is energetically favoured. In Section 5 we
summarize our results and conclusions.

\section{Nanosprings geometric features}  

A helical space curve is called {\it a curve of constant slope}, {\it
i. e.}, a curve whose tangent lines make a constant angle with a fixed
direction in the space (the helical axis)~\cite{dirk}. $\{{\bf n},{\bf
b}, {\bf t}\}$ is a frame, called {\it Frenet basis}, which is a
right-handed orthonormal basis defined at each point along a space
curve, where ${\bf t}$ is the tangent unit vector, ${\bf n}$ is the
{\it normal} unit vector and ${\bf b}$ is the {\it binormal} unit
vector. In order to define the {\it normal} and {\it binormal} vectors
we consider the plane defined by the points $P_1$, $P_2$ and $P_3$
belonging to the space curve. In the limit where $P_2$ and $P_3$
approach $P_1$, the plane is called the {\it osculating plane} of the
curve at $P_1$~\cite{dirk}. The tangent vector ${\bf t}$ belongs to
the osculating plane. ${\bf n}$ is defined as the unit vector
perpendicular to ${\bf t}$, that lies in the osculating plane while
${\bf b}$ is defined as the unit vector perpendicular to ${\bf t}$,
that is perpendicular to the osculating plane.

Let $I_1$ and $I_2$ be the principal moments of inertia of the
cross-section of a rod, along the two principal axes of the
cross-section. Along this paper, cross-sections of rods with $I_1=I_2$
will be called {\it symmetric cross-sections} and cross-sections of
rods with $I_1\neq I_2$ will be called {\it asymmetric
cross-sections}. According to this definition, circular and squared
cross-sections are symmetric, while elliptic and rectangular
cross-sections are asymmetric.

\begin{figure}[ht] 
  \begin{center}
  \includegraphics[height=90mm,width=75mm,clip]{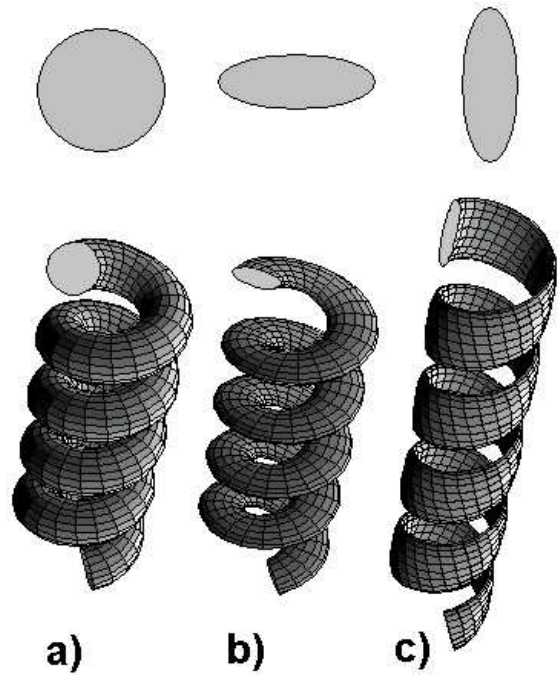} 
  \end{center}
  \caption{a) {\it Neutral} helix made of a rod with circular 
  cross section, b) {\it normal} and c) {\it binormal} helices 
  made of a rod with an asymmetric cross-section. The cross-section
  shape is depicted above the corresponding helix type.}
\label{fig1}
\end{figure}

Given a rod with asymmetric cross-section, we define the unit vector
${\bf d}$ lying in the cross-section plane along the direction of the
largest bending stiffness (it is the direction of the larger semiaxis
of an elliptic cross-section). The helical structure is said to be
{\it normal} ({\it binormal}) if ${\bf d}$ is in the direction of the
unit vector ${\bf n}$ (${\bf b}$). In the case of a rod with symmetric
cross-section, the {\it normal} and {\it binormal} structures
degenerate into one type of helix that we called a {\it neutral
helix}. Figure~\ref{fig1} displays examples of {\it neutral}, {\it
normal} and {\it binormal} helices, with the shape of their
corresponding cross-section.

Inspection of the transmission electron microscopy (TEM) images of the
amorphous nanosprings reported in Refs.~\cite{mc1,mc2,mc3,zhang} shows
that all of them are either {\it neutral} (Fig.~\ref{fig1}a) or {\it
binormal} (Fig.~\ref{fig1}c) helices, and none of them is a {\it
normal} helix (Fig.~\ref{fig1}b).

The silicon carbide (SiC) nanospring reported in Ref.~\cite{mc3} can
be classified as a neutral helix. According to Ref.~\cite{mc3}, this
nanospring is formed from a nanowire with circular cross-section. The
TEM image of the boron carbide (BC) nanospring depicted in panel b) of
Figure 10 of Ref.~\cite{mc1} is also a neutral helix grown from a
nanowire of circular cross-section.

The reported SiC~\cite{mc1}, BC~\cite{mc2}, and silicon oxide
(SiO$_2$)~\cite{zhang} nanosprings are clearly examples of {\it
binormal} helices. According to Ref.~\cite{zhang}, the SiO$_2$
nanospring was grown from a non-cylindrical nanowire. The reported SiC
and BC {\it binormal} nanosprings were formed from a nanowire with
rectangular cross-section~\cite{mc1,mc2}.

To our knowledge, the only {\it normal} helical nanostructure reported
in the literature is the crystalline ZnO nanohelix synthesized by Sb
induced thermal evaporation~\cite{gao}. The scanning electron
microscopy (SEM) images of these ZnO nanohelices show that each
helical period is formed by the sequence of six straight blocks, each
block growing in a given crystalline direction.

As up to now experimentalists have not reported the growth of
amorphous nanosprings of the {\it normal} type, we here present the
results of our investigation towards answering the question: {\it is
it possible to grow amorphous nanosprings of the normal type~?}
  
According to our previous analysis~\cite{douglas,fonseca1}, amorphous
nanosprings grown by the VLS mechanism are dynamically stable. This
stability stems from the intrinsic curvature produced by the catalytic
particle in the forming nanospring. The {\it intrinsic curvature} of a
rod represents its tridimensional shape when it is free from external
stresses. Goriely and Shipman have studied the dynamical stability of
{\it normal} and {\it binormal} helices~\cite{alain3} and showed that
intrinsically {\it normal} or {\it binormal} helical filaments are
always stable. Therefore, from the mechanical point of view, and in
agreement with our previous analysis~\cite{douglas,fonseca1}, there is
no mechanical prohibition for the existence of an amorphous {\it
normal} nanospring, thus both types of helical amorphous
nanostructures could be produced by the usual VLS mechanism. We have
conjectured that the shape of the liquid catalyst is the key to
explain the absence of normal amorphous nanospring. Thus we have
extended the VLS growth model to address also the case of
non-spherical liquid catalyst.


\section{The VLS model with non-spherical catalyst} 

\begin{figure}[ht] 
  \begin{center}
  \includegraphics[height=59mm,width=50mm,clip]{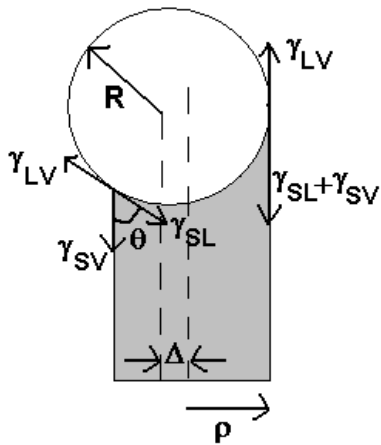}
  \end{center} \caption{Schematic diagram of the catalytic particle of
  radius $R$ atop a nanowire of radius $\rho$, whose center is shifted
  of $\Delta$ with respect to the axis of the nanowire.} 
\label{fig2}
\end{figure}

According to the VLS growth model, a liquid droplet of metal absorbs
the material from the surrounding vapor, and after super-saturation of
the absorbed material within the droplet, the excess material
precipitates at the liquid-solid interface forming the nanowire
beneath the metallic catalyst. The model is based on the interaction
between the surface tension of the liquid-vapor
(${\mathbf{\gamma}}_{LV}$), solid-vapor (${\mathbf{\gamma}}_{SV}$) and
solid-liquid (${\mathbf{\gamma}}_{SL}$) interfaces. McIlroy {\it et
al}~\cite{mc1,mc2} proposed that the helical growth process occurs due
to a contact angle anisotropy (CAA) at the catalyst-nanowire
interface. The trajectory of the metallic catalyst is driven by the
work needed to shear it from the surface of the nanowire. This work is
called the thermodynamic work of adhesion $W_A$ and can be computed in
terms of the surface tensions by~\cite{mc1}:
\begin{equation}
\label{wa}
\begin{array}{ccc}
W_A&=&\gamma_{SV}+\gamma_{SL}-\gamma_{LV} \; \\
&=&\gamma_{SV}(1-\cos\theta) 
\end{array}
\end{equation}
where $\theta$ is the angle between the surface tensions
${\mathbf{\gamma}}_{SL}$ and ${\mathbf{\gamma}}_{SV}$. Figure
\ref{fig2} reproduces the schematic diagram of a
spherical catalyst placed asymmetrically on the nanowire, in
accordance to the McIlroy {\it et al} modified VLS growth
model~\cite{mc1}.

Figures \ref{fig1}b) and \ref{fig1}c) display the cross-section of
{\it normal} and {\it binormal} helical structures, respectively. We
can see in these figures that, with respect to the plane of the page,
the cross-section of the {\it normal} helix has a horizontal dimension
larger than the vertical one, and vice-versa for the {\it binormal}
helix. The nanowire grows from the deposition of the material absorbed
by the liquid catalyst, so that, to grow structures with asymmetric
cross-section, the surface of contact between the catalytic particle
and the nanowire must follow the shape pattern of the cross-section.

An increase of the diameter of the spherical catalyst in the
anisotropic position with respect to the nanowire axis (as in
Fig.~\ref{fig2}), without increasing the nanowire diameter, increases
the asymmetry of the surface of contact between the catalyst and the
nanowire. However, according to McIlroy {\it et al}~\cite{mc1,mc2}, if
the diameter of the spherical catalyst increases systematically, so
that $\Delta/R$ decreases (see Fig.~\ref{fig2}), the CAA becomes less
significant and the work of adhesion becomes equal to that of the
symmetric configuration in which the nanowire grows linearly.

\begin{figure}[ht] 
  \begin{center}
  \includegraphics[height=68mm,width=100mm,clip]{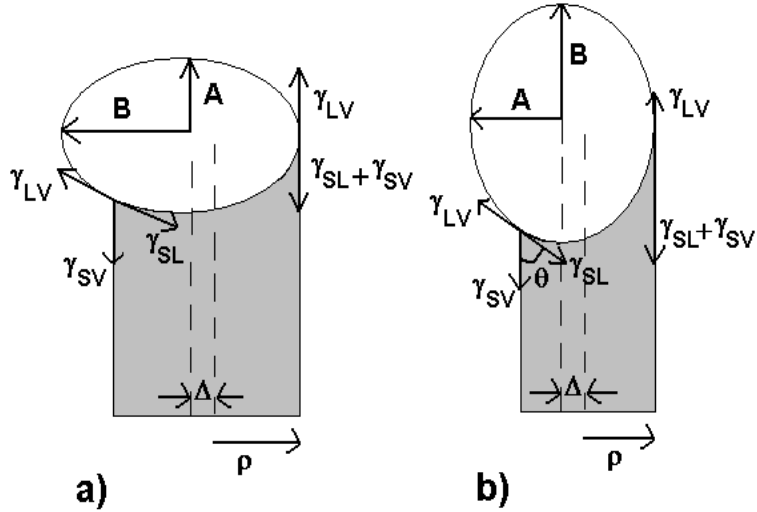}
  \end{center} 
  \caption{Schematic diagrams for the growth of a) {\it normal} and b)
    {\it binormal} nanosprings of elliptic cross-section. The center
    of the elliptic catalyst is shifted of $\Delta$ with respect to
    the axis of the nanowire. $\rho$ is the radius of the nanowire,
    and $A<B$ are the semiaxes of the elliptical catalyst in the plane
    of the figure.}
\label{fig3}
\end{figure}

Therefore, to have the surface of contact between the catalyst and the
nanowire following the pattern of a nanospring of asymmetric
cross-section, in the model considered by McIlroy {\it et
al}~\cite{mc1,mc2} (see Fig.~\ref{fig2}) we allow the catalyst to
possess a non-spherical shape. To produce a {\it normal} ({\it
binormal}) amorphous nanospring of elliptic cross-section, as
Fig.~\ref{fig1}b) (Fig~\ref{fig1}c)) we propose that the catalytic
particle is an ellipsoid as displayed in Fig.~\ref{fig3}a)
(Fig.~\ref{fig3}b)). The growth of an asymmetric nanospring, that is
driven by an elliptic catalyst, as shown in Fig.~\ref{fig3}, is
obtained in the same way as that driven by a spherical catalyst
(Fig.~\ref{fig2}): the growth rate velocity is larger at the interface
where the work of adhesion is smaller~\cite{mc1,mc2}. Of course, in
the case of growth of an asymmetric nanospring the shape of the
interface of contact between the catalyst and the nanowire is not
circular.

The TEM images of the transition regions from the linear nanowire
growth to nanospring growth (Figs.~15 and 17 of Ref.~\cite{mc1},
Fig.~2 of Ref.~\cite{mc2}, Fig.~5 of Ref.~\cite{mc3} and Fig.~3a of
Ref.~\cite{zhang}) give support to our proposal. All of these figures
display remnants of the catalyst inside the transition region. In
Ref.~\cite{mc1} these remnants are assumed to have spherical shape.
However, a minucious examination of these remnants shows that some of
them are not spherical. Fig.~3a of Ref.~\cite{zhang} presents an
oval-shaped particle inside the transition region. Fig.~15b of
Ref.~\cite{mc1} or Fig.~2b of Ref.~\cite{mc2} shows that the part of
the catalyst lying inside the nanowire in the transition region is
approximatelly elliptic. The shape of the particle inside the
transition from linear to helical SiC nanowire displayed in Fig.~5 of
Ref.~\cite{mc3} has a more complex shape.

According to McIlroy {\it et al}~\cite{mc1}, changes in the materials
absorbed by the catalyst, the introduction of additional elements, and
changes in the local temperature can introduce imbalances in the
energy related to the surface tension of the liquid-vapor
(${\mathbf{\gamma}}_{LV}$), solid-vapor (${\mathbf{\gamma}}_{SV}$) and
solid-liquid (${\mathbf{\gamma}}_{SL}$) interfaces, that lead to
variations in the shape and mass of the catalyst during the growth
process of the nanowire.

Therefore, we propose that after the transition from linear to helical
growth of the nanowire, the ejected part of the catalyst has to
possess an asymmetric shape to drive the growth of a nanospring with
asymmetric cross-section.

\section{Why amorphous {\textit{normal}} nanosprings have not been
  observed ?} 

The idea of considering non-spherical catalyst is essential for
explaining the formation of a nanospring with asymmetric
cross-section. Now, to explain why amorphous {\it normal} nanosprings
have not been synthesized we have to look into the Contact Angle
Anisotropy (CAA)~\cite{mc1,mc2}. According to McIlroy {\it et al}
proposal~\cite{mc1,mc2}, the CAA has to be significant for the helical
growth of a nanowire. So, if the CAA is significant for the growth of
{\it normal} and {\it binormal} nanosprings, then both types of
nanosprings could be grown through the VLS mechanism. To explain the
absence of amorphous {\it normal} nanosprings, we shall analyse the
significance of the CAA for growing {\it normal} (Fig.~\ref{fig3}a))
and {\it binormal} (Fig.~\ref{fig3}b)) nanosprings.

According to McIlroy {\it et al}~\cite{mc1,mc2}, if the diameter of
the spherical catalyst increases systematically, so that $\Delta/R$
decreases (see Fig.~\ref{fig2}), the CAA becomes less significant and
the work of adhesion becomes equal to that of the symmetric
configuration in which the nanowire grows linearly. As we are dealing
with non-spherical catalyst we propose that the significance of the
CAA comes from the ratio $\Delta/X$, where $X$ is the dimension of the
catalyst particle along the direction of the shift of the catalyst
with respect to the nanowire axis. In the case of a spherical catalyst
of radius $R$, $X=R$, giving McIlroy {\it et al}~\cite{mc1,mc2} ratio
$\Delta/R$, while in the case of an elliptical catalyst $X=B$ ($X=A$)
in the scheme depicted in Fig.~\ref{fig3}a) (Fig.~\ref{fig3}b)).

We shall analyze the significance of the CAA for both cases depicted
in Fig.~\ref{fig3}. The magnitude of $B$ ($A$) will determine the
significance of the CAA for the scheme depicted in Fig.~\ref{fig3}a)
(Fig.~\ref{fig3}b)). Since $\Delta/B$ is always smaller than
$\Delta/A$, we expect that the scheme displayed in Fig.~\ref{fig3}a)
is less favourable for growing a helical nanowire than that of
Fig.~\ref{fig3}b). To show this, we have calculated the work of
adhesion along the interface solid-liquid-vapor of an elliptical
metallic catalyst in contact with the nanowire, for the two situations
depicted in Fig.~\ref{fig3}.

\begin{figure}[ht] 
  \begin{center}
  \includegraphics[height=60mm,width=60mm,clip]{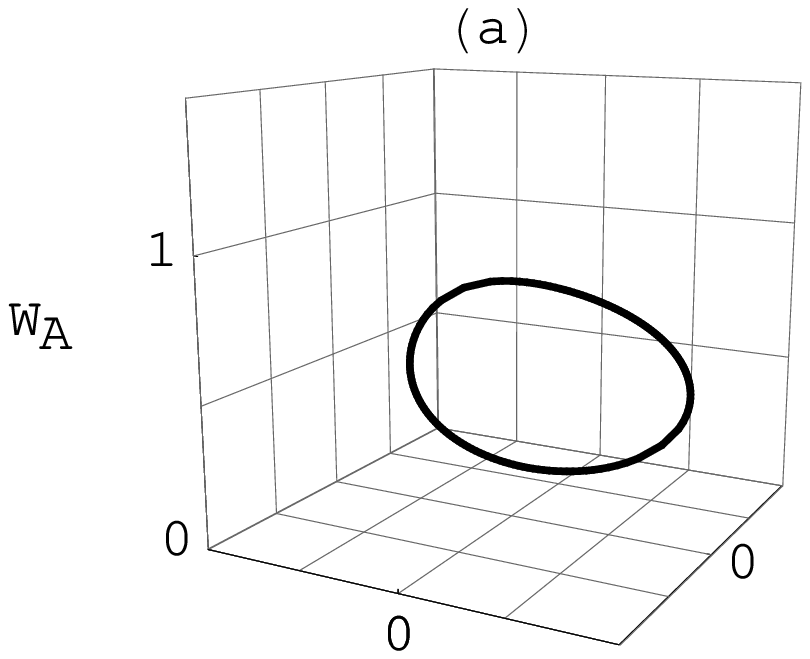} 
  \includegraphics[height=60mm,width=60mm,clip]{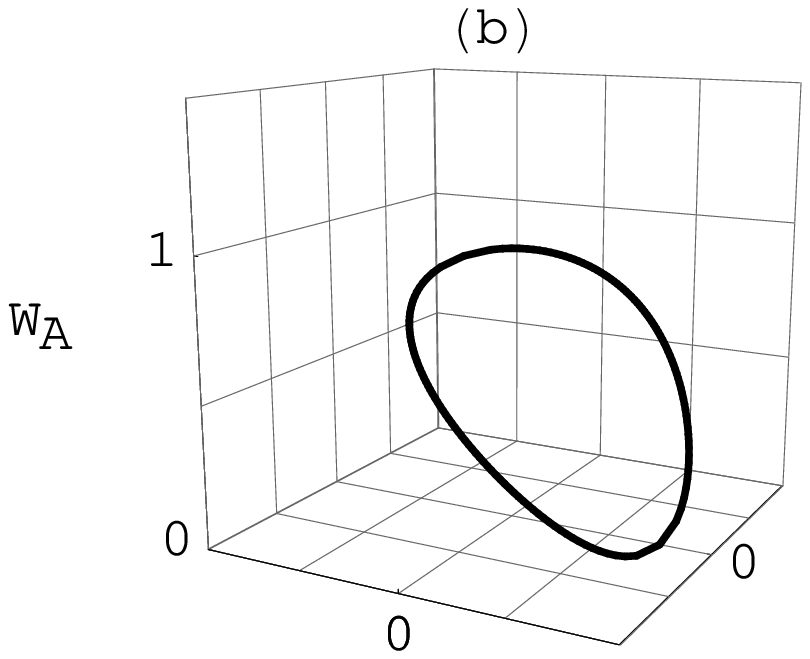} 
  \end{center}
  \caption{The work of adhesion $W_A$ (in arbitrary units) along the
    interface solid-liquid-vapor formed by an elliptic catalytic
    particle ($A=1.5,B=2.0$) onto a nanowire of $\rho=1.0$, with
    $\Delta=0.5$, for (a) {\it normal} growth, depicted in Fig.
    \ref{fig3}a), and (b) {\it binormal} growth, depicted in Fig.
    \ref{fig3}b).}
\label{fig4}
\end{figure}

Fig.~\ref{fig4} displays the work of adhesion $W_A$ for growing a
helical nanostructure, driven by an elliptical catalyst having
dimensions $A=1.5$, $B=2.0$, onto a nanowire with $\rho=1.0$, and
$\Delta=0.5$ (see Fig.~\ref{fig3}). The condition for the optimal
geometry to promoting helical growth has been inferred from that
obtained by McIlroy {\it et al}~\cite{mc2} for the growth of a helical
nanowire in the case of a spherical catalyst.  They found that
$R/\rho\simeq1.5$~\cite{mc1,mc2}.  As the catalyst here has elliptic
shape, we propose to replace $R$ by $X$, so that the condition for the
optimal geometry to promoting helical growth can be written as
$X/\rho\simeq 1.5$. The chosen values of $\rho$ and $\Delta$ are such
that the significance of the CAA gives the optimal geometry for the
promotion of helical growth in the case of spherical
catalyst~\cite{mc2} with radius $R=1.5$. Panel (a) of Fig.~\ref{fig4}
shows the work of adhesion for the scheme of Fig.~\ref{fig3}a) for
growing a {\it normal} nanospring ($(X/\rho)=(B/\rho)=2.0$).  Panel
(b) of Fig.~\ref{fig4} shows the work of adhesion for the scheme of
Fig.~\ref{fig3}b) for growing a {\it binormal} nanospring
($(X/\rho)=(A/\rho)=1.5$).  The difference between the minimum and the
maximum values of the work of adhesion, hereafter called $d_{WA}$,
gives a measure of the amount of anisotropy in the contact angle and,
therefore, the CAA significance for growing a helical nanostructure in
that situation.  $d_{WA}=0.458$ for the case displayed in the
Fig.~\ref{fig4}a) while $d_{WA}=0.922$ for the case displayed in the
Fig.~\ref{fig4}b), thus indicating that for the elliptic catalytic
particle having dimensions $A=1.5, B=2.0$, the optimal geometry to
promoting helical growth corresponds to the {\it binormal} nanohelix
for which $(X/\rho)=(A/\rho)=1.5$.

To show that the larger the value of $\Delta/X$, the smaller the
significance of the CAA in the helical growth, we have calculated
$d_{WA}$ keeping $B$ fixed and varying $A$, using the parameters of
Fig.~\ref{fig4} ($\rho=1.0$, $\Delta=0.5$). We have considered three
fixed values of $B$: (i) $B=2.0$ and $B=3.0$ for the growth scheme in
Fig.~\ref{fig3}b); (ii) $B=1.5$ for the growth scheme in
Fig.~\ref{fig3}a). To consider only catalysts for which their extreme
edges coincide with the extreme edge of the nanowire, as considered by
McIlroy {\it et al}~\cite{mc2} (see Fig. \ref{fig2} or \ref{fig3}), we
only calculate the $d_{WA}$ for $X\ge(\rho+\Delta)$ (in this case
$\rho+\Delta=1.5$). So, in the case of the growth scheme in
Fig.~\ref{fig3}b), for which $X=A$, we have varied $A \in [1.5,9.0]$
for both values of fixed $B$. In the case of the growth scheme in
Fig.~\ref{fig3}a) (for which $X=B$) we have fixed $B=1.5$, and varied
$A \in [0.75,3.30]$.

\begin{figure}[ht] 
  \begin{center}
  \includegraphics[height=46mm,width=65mm,clip]{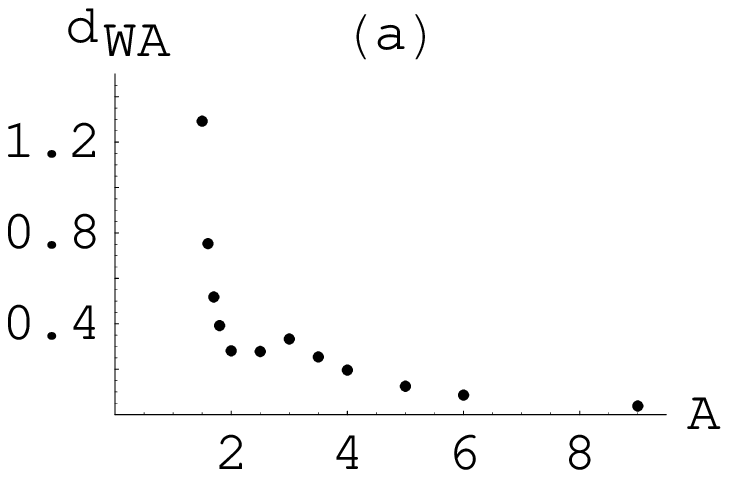} 
  \includegraphics[height=46mm,width=65mm,clip]{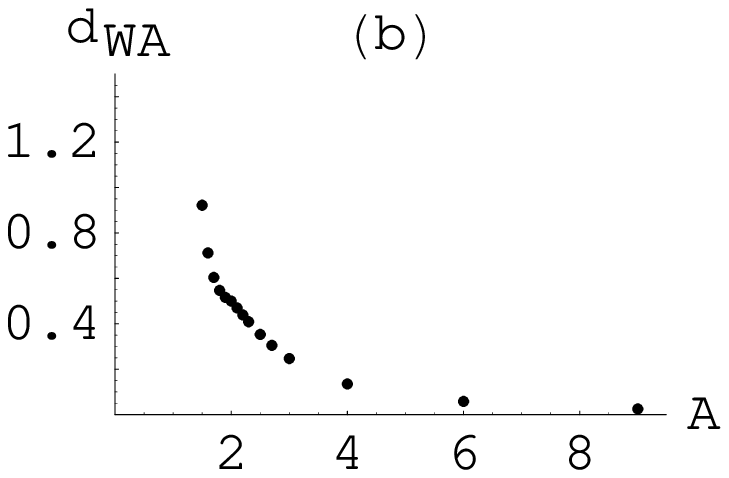} 
\end{center}
  \caption{The difference, $d_{WA}$ (in arbitrary units), between the
    minimum and the maximum values of the work of adhesion as a
    function of $A \in [1.5,9.0]$, for the scheme depicted in Fig.
    \ref{fig3}b), with $\rho=1.0$ and $\Delta=0.5$. $B$ is fixed: a)
    $B=3.0$; b) $B=2.0$ .}
\label{fig5}
\end{figure}

\begin{figure}[ht] 
  \begin{center}
  \includegraphics[height=66mm,width=85mm,clip]{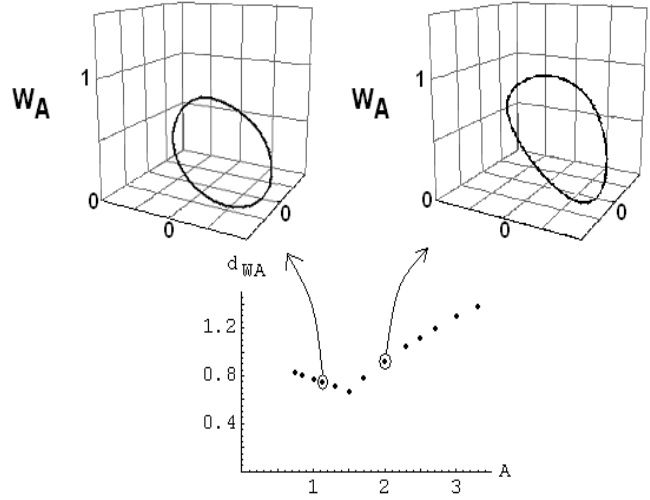} 
  \end{center}
  \caption{Bottom: the difference, $d_{WA}$ (in arbitrary units),
    between the minimum and the maximum values of the work of adhesion
    as a function of $A \in [0.75, 3.30]$, for $B=1.5$ fixed,
    $\rho=1.0$ and $\Delta=0.5$, for the scheme depicted in Fig.
    \ref{fig3}a). Notice that the nanohelix is {\it normal} ({\it
      binormal}) for $A<1.5$ ($A>1.5$). Top left (top right) is the
    work of adhesion, in arbitrary units, along the interface
    solid-liquid-vapor formed by the catalytic particle, with $A=1.13$
    ($A=2.00$), onto the nanowire with the parameters above (they
    correspond to the points encircled in the $d_{WA} \times A$
    plotting).}
\label{fig6}
\end{figure}

The $d_{WA} \times A$ plotting for the cases $B=3.0$, $B=2.0$, and
growth scheme in Fig.~\ref{fig3}b), are displayed in
Figs.~\ref{fig5}a) and \ref{fig5}b), respectively.  Notice that for
$A<B$ ($A>B$) the nanohelix is {\it binormal} ({\it normal}) and it is
neutral for $A=B$ (case of spherical catalyst). Fig.~\ref{fig5} shows
that $d_{WA}$ exhibits a small peak at $A=B$, but for all other values
of $A$ it decreases as $A$ increases, implying that the significance
of CAA decreases as $\Delta/A$ decreases. The significance of CAA is
larger for $A=1.5$ therefore it corresponds to the optimal geometry to
promoting helical growth corroborating our proposal of replacing the
condition $R/\rho \simeq 1.5$ by $X/\rho \simeq 1.5$ ($X=A$ for the
growth scheme in Fig.~\ref{fig3}b)).
The $d_{WA} \times A$ plotting for the case $B=1.5$, and growth scheme
in Fig.~\ref{fig3}a), is displayed in Fig.~\ref{fig6} (bottom), for $A
\in [0.75,3.30]$. In this case $X=B$, and according to our proposal
the condition for the optimal geometry to promoting helical growth is
always satisfied since $(X/\rho)= (B/\rho)= 1.5$.  The case of
spherical catalyst, $A=B=1.5$, corresponds to the optimal geometry to
promoting helical growth according to McIlroy {\it et al}~\cite{mc2}.
Notice that for $A<B=1.5$ ($A>B=1.5$) the nanohelix is {\it normal}
({\it binormal}), and it is neutral for $A=1.5$ (this corresponds to
the case of spherical catalyst).  Fig.~\ref{fig6} (bottom) shows that
$d_{WA}$ is decreasing for $A<B$, while for $A>B$ it is increasing (at
a faster rate). At the top of Fig.~\ref{fig6} we display the work of
adhesion along the interface solid-liquid-vapor formed by the elliptic
catalytic particle onto the nanowire for the cases $A=1.13$ (top left)
and $A=2.00$ (top right), the latter case being the same one displayed
in Fig.~\ref{fig4}b).  $d_{WA}= 0.922$ for the case $A=2.00$
(Fig.~\ref{fig6}, top right), and $d_{WA}=0.743$ for the case $A=1.13$
(Fig.~\ref{fig6}, top left).  In these two cases the elliptic catalyst
has its major axis approximately 1.33 times its minor axis, and our
results indicate that the {\it binormal} type is favoured
energetically.

From the $d_{WA} \times A$ plottings in Figs.~\ref{fig5} and
\ref{fig6}, we conclude that the growth of a {\it binormal nanospring}
is favoured energetically whenever both semiaxes of the elliptic
catalyst are $\ge(\rho+\Delta)$. If the major semiaxis is $\ge(\rho +
\Delta)$ while the minor semiaxis is $<(\rho+\Delta)$, then it is not
possible to grow a {\it binormal nanospring}, and in this case it
would be possible to grow a {\it normal nanospring} as indicated by
the $d_{WA} \times A$ plotting in Fig.~\ref{fig6} when $A<1.5$. So,
depending on the dimensions of the elliptic catalyst relative to
$(\rho+\Delta)$ the CAA may be significant for growing a {\it normal}
nanohelix whenever $A<(\rho+\Delta)$ (see Fig.~\ref{fig3}).

In the particular case of $B\simeq A$, the grown helix could be either
{\it normal} or {\it binormal} as shown in the $d_{WA} \times A$
plottings (Figs.~\ref{fig5} and \ref{fig6}) within the region around
$B=A$.  However, in this case, the shape of the catalyst is
approximately spherical, and the resulting helical structure will be
very similar to that of a {\it neutral} helix (spherical catalyst).
Therefore, {\it normal} or {\it binormal} nanosprings grown by an
almost spherical catalyst are not experimentally distinguible from a
{\it neutral} nanospring.

The TEM images of the transition regions from the linear nanowire
growth to nanospring growth reported in the literature (Figs.~15 and
17 of Ref.~\cite{mc1}, Fig.~2 of Ref.~\cite{mc2}, Fig.~5 of
Ref.~\cite{mc3} and Fig.~3a of Ref.~\cite{zhang}), show that the
remnant frozen part of the catalyst is not spherical. This fact,
together with our results in Figs.~\ref{fig5} and \ref{fig6}, allows
us to infer that the part of the catalyst that was ejected, and drove
the helical growth, had the elliptic shape displayed in
Fig.~\ref{fig3}b).

The formation of helical nanowires of rectangular cross-section, as
the amorphous BC nanospring displayed in Figure~1 of Ref.~\cite{mc2}
or the amorphous SiC nanospring displayed in Figure~13 of
Ref.~\cite{mc1}, can be explained using our extended VLS growth model
for a rectangular metallic catalyst. If the smaller side of the
rectangular catalyst is not larger than the displacement $\Delta$, the
CAA is significant and the helical growth can occur.


\section{Conclusions}

We have studied the geometric features of several types of nanosprings
reported in the literature. The published images of several
nanosprings and nanohelices were analyzed and we have verified the
non-existence of one type of helical structure in the case of
amorphous nanostructures: {\it normal nanohelix}. In the case of
amorphous materials, we discussed the importance of the shape of the
catalyst in order to drive the growth of a nanospring of asymmetric
cross-section. We extended the modified VLS growth model to include
non-spherical shapes of the catalyst so as to explain the growth of
asymmetric amorphous nanosprings. The conformation of the amorphous
nanosprings seen in the TEM images are explained by our proposal.

We have shown that the non-spherical shape of the metallic catalyst,
within the model proposed by McIlroy {\it et al}~\cite{mc1,mc2}, can
induce the growth of amorphous nanosprings with asymmetric
cross-section. We have also shown that the anisotropy in the work of
adhesion along the interface liquid-solid is more significant for
growing a {\it binormal nanohelix} than for growing a {\it normal
nanohelix}, thus explaining the absence of amorphous {\it normal}
nanosprings.

From the present study we conclude that the resulting type of helical
nanostructure (its cross-section) is related to the shape of the
metallic catalyst that induced its growth. So, from the type and shape
of the nanospring it is possible to qualitatively infer the shape of
the metallic catalyst.  For example, if the period of the turns
changes along the nanospring, as seen in the SiO$_2$ nanosprings of
Ref.~\cite{zhang}, our analysis suggests that the size and shape of
the catalyst must have changed during the nanospring formation.

Our results are in perfect agreement with the experimental TEM images
of various nanosprings and provide new insight on the geometric and
mechanical characteristics of both types of helices. It is well
established that for some growth phenomena at nanoscale the presence
of the catalytic particles is fundamental, nevertheless the details of
how they define the nanostructure morphology is not well understood.
In the present work, we show how the catalytic particle shape is
important to determine the morphological symmetries. Our study shows
that when both semiaxes of the elliptic catalyst are
$\ge(\rho+\Delta)$ the growth of amorphous {\it binormal} nanospring
is energetically favoured through the VLS growing model. So, for
$\Delta=0.5$ and $\rho=1.0$, and the elliptic catalytic particle with
semiaxes 1.5 and 2.0, Fig.~\ref{fig4} shows that the {\it binormal}
nanohelix is clearly favoured energetically.  It might be possible to
grow an amorphous {\it normal} nanospring within the VLS model only if
the elliptic catalytic particle has its minor semiaxis
$<(\rho+\Delta)$. We hope that our analysis will stimulate further
theoretical and experimental investigations for growing the various
types of helical nanostructures.

This work was partially supported by the FAPESP, CNPq, IMMP, IN,
THEO-NANO and FINEP. AFF acknowledges a scholarship from the Brazilian
Agency CNPq.



\end{document}